\documentclass[reprint,showpacs,amsmath,amssymb,aps,pra,superscriptaddress]{revtex4-1}

\usepackage[ascii]{inputenc}
\usepackage[T1]{fontenc}
\usepackage[english]{babel}
\usepackage{graphicx}
\usepackage{microtype}
\usepackage[pdftex,colorlinks]{hyperref}		 
\usepackage{xcolor}
\hypersetup{%
  linkcolor=blue,%
  citecolor=blue,%
  urlcolor=blue,%
  pdftitle={PT-symmetric waveguides revealing third-order exceptional points},%
  pdfauthor={Jan Schnabel, Holger Cartarius, J\"org Main, G\"unter Wunner, Walter Dieter Heiss}%
}
\usepackage{siunitx}
\usepackage{braket}

\renewcommand{\Re}{{\rm Re}\,}	
\renewcommand{\Im}{{\rm Im}\,}	

\newcommand\ee{\mathrm{e}}
\newcommand\ii{\mathrm{i}}
\newcommand{\PT}{$\mathcal{PT}$}

\begin{document}

\title{\PT-symmetric wave guide system with evidence of a third-order exceptional point}

\author{Jan Schnabel}
\email[]{jan.schnabel@itp1.uni-stuttgart.de} 
\author{Holger Cartarius}
\author{J\"org Main}
\author{G\"unter Wunner}
\affiliation{1. Institut f\"ur Theoretische Physik, 
  Universit\"at Stuttgart, 70550 Stuttgart, Germany}
\author{Walter Dieter Heiss}
\affiliation{Department of Physics, University of Stellenbosch,
  7602 Matieland, South Africa}
\affiliation{National Institute for Theoretical Physics (NITheP), Western
  Cape, South Africa}

\date{\today}

\begin{abstract}
  An experimental setup of three coupled \PT-symmetric wave guides showing
  the characteristics of a third-order exceptional point (EP3) has been
  investigated in an idealized model of three delta-functions wave guides
  in W.~D. Heiss and G.~Wunner, J. Phys. A \textbf{49}, 495303 (2016). Here
  we extend these investigations to realistic, extended wave guide systems. 
  We place major focus on the strong parameter sensitivity rendering the
  discovery of an EP3 a challenging task. We also investigate the vicinity of
  the EP3 for further branch points of either cubic or square root type
  behavior.
\end{abstract}

\maketitle

\section{Introduction}
\label{sec:introduction}
The term ``exceptional point'' (EP) originates from a purely
mathematical context and describes branch point singularities in the
spectrum of parameter-dependent linear operators~\cite{Kato}. However,
by now there is an overwhelming interest in physics~\cite{Heiss12} on
this topic both theoretically (see e.g.~\cite{Heiss90,Heiss2000,%
  Hernandez2006,Lefebvre2009,Cartarius09,Cartarius2011b,Gutoehrlein13,
  Wiersig2014a,Schwarz2015a,Menke2016a}) and  experimentally (see,
e.g.~\cite{Philipp2000,Dembowski2003,Dietz2007,Stehmann2004,Lawrence2014a,%
  Gao2015a,Doppler2016a,Xu2016a}). In general EPs are positions in some
parameter space, at which two (EP2) or even $N>2$ (EP$N$) eigenvalues, as well
as the corresponding eigenvectors, coalesce in a branch point singularity.
These points can be found in the vicinity of level repulsion if one external
system parameter is analytically continued into the complex
plane~\cite{Heiss99}. This renders the underlying Hamiltonian describing
the physics of the system to be no longer Hermitian \cite{Morse1953}. In fact,
exceptional points can only occur for non-Hermitian Hamiltonians. The
manifestation of exceptional points is not only restricted to quantum systems.
For non-Hermitian systems, they also occur in classical mechanics
\cite{HeissWunner15} as well as in optics~\cite{Lee09,Longhi10,Guo09,%
Chong2011a,Ge2014a,Chang2014,Feng2014a,Hodaei2014a} and microwave
cavities~\cite{Dembowski01}. In order to obtain a unitary theory the
non-Hermiticity requires the definition of a new inner product -- the
bi-orthogonal product or \emph{c-product}~\cite{Brody14,Moiseyev}.
At the exceptional points the corresponding Hilbert space becomes defective
in that the number of eigenvectors is reduced as a consequence of the
coalescence.

EPs show more characteristic properties than those mentioned above: In their
simplest manifestation, i.e., for an EP2, the two eigenvalues can be
mathematically described by two branches of the same analytic function, thus
showing typical square root behavior. This means, if one encircles the EP2
along a closed loop in the physical parameter space the corresponding
eigenvalues forming an EP2 permute. Exceptional points of higher order, e.g.,
third-order exceptional points, show cubic root behavior, i.e., one
\emph{typically} observes a threefold state exchange performing a closed loop
around the EP3. Moreover, since also the eigenvectors coalesce at the
exceptional point -- forming a \emph{self-orthogonal} state~\cite{Moiseyev}
-- the corresponding Hamiltonian in matrix representation is no longer
diagonalizable. With a similarity transformation, however, one can transform
it to a \emph{Jordan normal form}. There an exceptional point of order $N$ is
represented in terms of an $N$-dimensional Jordan block~\cite{Guenther07}.

Exceptional points appear in particular in \PT-symmetric systems,
i.e., systems which are symmetric under the combined action of the parity
operator $\mathcal{P}$ and the time reversal operator $\mathcal{T}$. Bender and
Boettcher~\cite{Bender98} demonstrated that \PT-symmetric
non-Hermitian Hamiltonians can possess real eigenvalues. When the real
eigenvalues coalesce and turn into complex conjugates the underlying
\PT\ symmetry is broken. The parameter set at which the symmetry is broken marks
the position of an exceptional point. As this class of non-Hermitian
Hamiltonians is in particular predestined for the occurrence of EPs they have
been investigated in a wide range of systems ranging from fundamental questions
in quantum mechanics \cite{Bender1999,Znojil1999,Jones2010}, quantum field
theories \cite{Bender2012,Mannheim2013} to Bose-Einstein condensates in the
mean-field approximation \cite{Graefe2012a,Dast13,Abt2015a,%
  GutoehrleinSchnabel15} and many-particle descriptions \cite{Graefe08two,%
  Dast2016a}, where complex potentials model the gain and loss of particles
\cite{Kreibich16,Dast14a}. $\mathcal{PT}$ symmetry has also been studied in
cavities for electromagnetic waves \cite{Bittner2014a,Peng2014a,Bender13},
optical structures with complex refractive indices \cite{Makris2008,%
  El-Ganainy2007}, and in electronic devices \cite{Schindler2011}. Spectral
singularities in $\mathcal{PT}$-symmetric potentials \cite{Mostafazadeh2009a}
turned out to be connected with the amplification of waves
\cite{Mostafazadeh2013a} and the lasing threshold \cite{Mostafazadeh2013b}.

Klaiman \emph{et al.}~\cite{Klaiman08} proposed an experimental setup
of two coupled \PT-symmetric wave guides with complex refractive index
for the visualization of second-order branch points. The imaginary
parts are interpreted as gain (loss) of the field intensity, e.g., by optical
pumping and absorption. Its strength controls the non-Hermiticity. Their
investigations showed the coalescence of the system's eigenmodes,
experimentally observable in terms of an increasing beat length in the power
distribution of the total field. The predictions received convincing
experimental confirmation 2010 by R\"uter \emph{et al.}~\cite{Rueter10}.

While the physics of EP2s is well investigated, lesser attention has
hitherto been paid to exceptional points of higher order~\cite{Graefe08, 
HeissChiral,Graefe12,HeissWunner16,Ge2015a,Ding2016a,Zin2016a,Jing2016a}. New
effects were shown in the different theoretical models of higher order EPs.
In~\cite{HeissChiral} a chiral behavior of the eigenfunctions in the
neighborhood of three coalescing eigenfunctions was
reported. In~\cite{Graefe12} it was shown that encircling an EP3 does not
necessarily show the typical third-root behavior. In this context a possible
experiment made up of three coupled wave guides was proposed in terms of an
abstract mathematical matrix model. Here our work sets in. Encouraged by the
experimental confirmation of the wave guide system investigated
in~\cite{Klaiman08} we extend this model by placing a third wave guide between
those with gain and loss but with only a real part of the refractive index
that may be different from that of the outer ones. We show that this model
gives rise to a third-order exceptional point by solving the whole system
semi-analytically. We work out explicitly the appearance of further EP2s or
EP3s in the vicinity of the original EP3 as was discussed qualitatively in
\cite{HeissChiral,Graefe12}.

The paper is organized as follows. Sec.~\ref{sec:system} introduces
the system including the corresponding equations. These are solved in
Sec.~\ref{sec:results}, where we demonstrate the manifestation of the EP3,
its verification as well as the total power distribution. In
Sec.~\ref{sec:EP2s} we explicitly demonstrate the 
additional EP2s and EP3s in the space of the system's physical
parameters. In Sec.~\ref{sec:summary} we summarize the crucial points and give
an outlook to ongoing work.

\section{The \PT-symmetric optical wave guide system}
\label{sec:system}

\begin{figure}
  \centering
  \includegraphics[width=\linewidth]{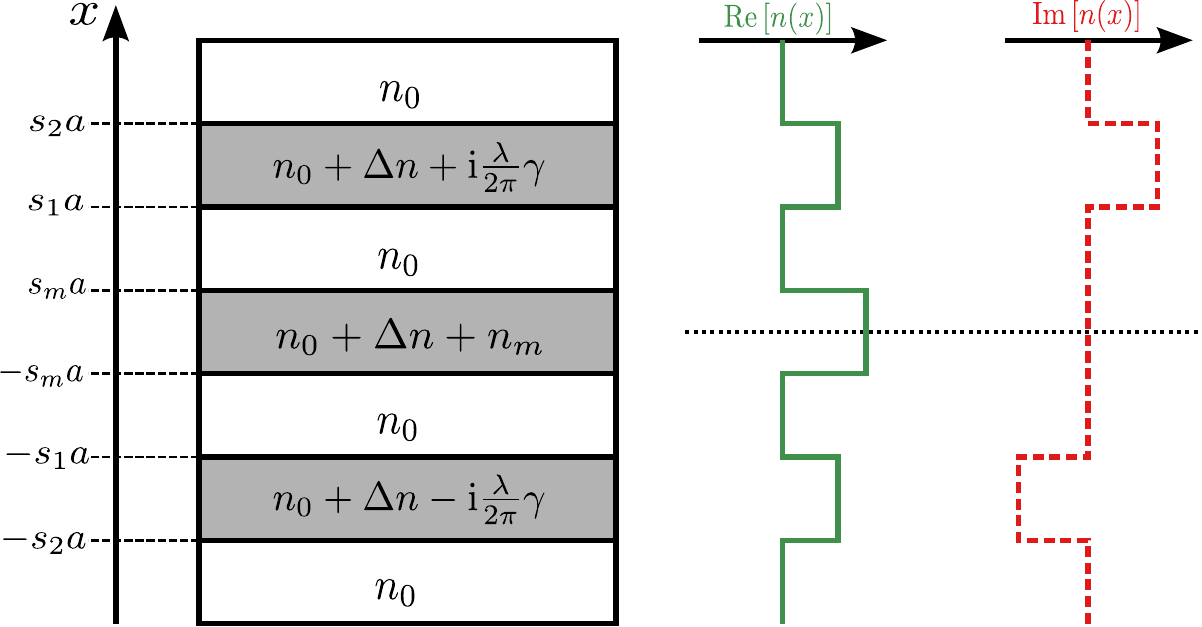}
  \caption{\PT-symmetric directional coupling. The structure consists
    of three coupled slab wave guides on a background material with
    index $n_0=3.3$. The right hand side displays the real and
    imaginary part of the refractive index, which only vary in $x$
    direction. Using the dimensionless parameters $s_m$ and $s_{1,2}$
    the wave guides' width as well as the separation
    between them is adjusted. Here a value of $a=\SI{2.5}{\micro\m}$ is
    chosen. Also note that we allow for a larger real index difference
    between the middle wave guide and the background material as
    compared to the outer ones by adding an additional term $n_m$ to
    the fixed one $\Delta n = 1.3\times 10^{-3}$. The imaginary part of
    the index can be controlled by the gain-loss coefficient $\gamma$
    with the vacuum wavelength taken to be
    $\lambda=\SI{1.55}{\micro\m}$.}
  \label{fig:system}
\end{figure}

We model a \PT-symmetric wave guide system for the experimental
observation of a third-order branch point with three coupled planar
wave guides on a background material with refractive index $n_0 =
3.3$, as depicted in Fig.~\ref{fig:system}.
We assume the refractive index to vary only in $x$ direction with a symmetric
index guiding profile and an antisymmetric gain-loss profile,
i.e., $n(x)=n^*(-x)$  to sustain \PT\ symmetry. For three wave guides we
basically follow the approach used in \cite{Klaiman08} for two wave guides,
but in addition we allow for more flexibility of the wave guides' parameters.
Their width and the separation between them can be varied with dimensionless
scaling factors $s_m$ and $s_{1,2}$ in order to define distances via the
constant length scale $a=\SI{2.5}{\micro\m}$ (cf. Fig.~\ref{fig:system}).
Moreover we chose $\Delta n = 1.3\times 10^{-3}$ and allow for a different real 
index difference between the middle wave guide and the background material as
compared to the outer ones by adding an additional term $n_m$. The imaginary
part of the refractive index can be controlled by the non-Hermiticity parameter
$\gamma$ with the vacuum wavelength taken to be
$\lambda=\SI{1.55}{\micro\m}$.
A realization of the system studied in this work is possible with GaAs
or ZnSiAs$_2$ as guiding material. The variations of the refractive index
are possible with a carrier-induced change \cite{Dutta1992a}, electric field
induced changes \cite{Susa1995a}, or femtosecond-scale switching
\cite{Ganikhanov1999a}.

The direction of propagation in the wave guides is taken to be the $z$
axis, such that the wave equation for the transverse-electric modes reads
\begin{equation}
  \label{eq:wave-equation}
  \left(\frac{\partial^2}{\partial x^2} +
    k^2n(x)^2\right)\mathcal{E}_y(x) = \beta^2\mathcal{E}_y(x)\,,
\end{equation}
where the $y$ component of the electric field is given by
$E_y(x,z,t) = \mathcal{E}_y(x)\ee^{\ii (\omega t - \beta z)}, $
with $k=2\pi/\lambda$ and the propagation constant $\beta$. Obviously
Eq.~\eqref{eq:wave-equation} is formally equivalent to a one-dimensional 
stationary Schr\"odinger equation with potential $V(x) =
-\frac{1}{2}k^2n(x)^2$ and energy eigenvalue $E = -\frac{1}{2}\beta^2\,.$ 
Thus the quantum mechanical analogue of the arrangement shown in
Fig.~\ref{fig:system} is a configuration of three finite potential
wells with gain or loss in the two outer wells.

 Because of the underlying \PT\ symmetry there is some range of
$\gamma$, for which $\beta$ is purely real. The point at which all three modes
break this symmetry simultaneously and become complex, is associated with an
EP3. The challenging part in a numerical simulation, as well as in an
experiment, is to find the correct values for the system parameters
$(\beta,\gamma,s_m,s_1,s_2,n_m,\Delta n)$ to determine this point.

\section{Solution of the full wave guide system}
\label{sec:results}

\subsection{Semi-analytical approach and method for finding an EP3}

The stationary modes can be taken to be
\begin{align}
  \label{eq:eigenmodes}
  \tilde{\mathcal{E}}_y(x) &= 
  \begin{cases}
      A_1\ee^{\kappa x} + A_2\ee^{-\kappa x} &:\; -\infty < x < -s_2a \\
      B\ee^{\ii q_lx} + C\ee^{-\ii q_l x} &:\; -s_2a\leq x\leq -s_1a \\
      D_1\ee^{\kappa x} + D_2\ee^{-\kappa x} &:\; -s_1a < x < -s_ma \\
      F\ee^{\ii q_m x} + G\ee^{-\ii q_m x} &:\; -s_ma \leq x\leq s_ma \\
      H_1\ee^{\kappa x} + H_2\ee^{-\kappa x} &:\; s_ma < x < s_1a \\
      K\ee^{\ii q_r x} + L\ee^{-\ii q_rx} &:\; s_1a\leq x \leq s_2a \\
      M_1\ee^{-\kappa x} + M_2\ee^{\kappa x} &:\; s_2a < x < \infty
    \end{cases}
\end{align}
with the parameters
\begin{subequations}
  \begin{align}
    \kappa^2 &= \beta^2 - k^2n_0^2\,,\label{eq:wavefuncs-params-a}\\
    q_l^2 &= -\beta^2 + k^2\left(n_0+\Delta n - \ii\frac{\lambda}{2\pi}\gamma\right)^2\,,\label{eq:wavefuncs-params-b}\\
    q_m^2 &= -\beta^2 + k^2\left(n_0+\Delta n + n_m\right)^2\,,\label{eq:wavefuncs-params-c}\\
    q_r^2 &= -\beta^2 + k^2\left(n_0+\Delta n + \ii\frac{\lambda}{2\pi}\gamma\right)^2\,.\label{eq:wavefuncs-params-d}
  \end{align}
\end{subequations}
Similar to the procedure in \cite{Mostafazadeh2009a} the continuity conditions
at the potential barriers can be combined in a \emph{transition matrix}
$\mathbf{T}\in\mathbb{C}^{2\times 2}$ relating the  coefficients of the two
outermost parts of the system \cite{Yeh1988}. Thus the whole physics of the
system is incorporated in this matrix. To obtain physical meaningful solutions
out of Eq.~\eqref{eq:eigenmodes} the condition
\begin{equation}
  \label{eq:conditions}
  A_2 = M_2 = 0
\end{equation}
has to be fulfilled. Then the relation just mentioned between the system's
left- and right-hand sides reads
\begin{align}
  \label{eq:transition-matrix}
    \begin{pmatrix}
      A_1\\
      0
    \end{pmatrix}
  &= \mathbf{T}\cdot
    \begin{pmatrix}
      M_1\\
      0
    \end{pmatrix}\nonumber \\
  &=
  \begin{pmatrix}
      T_{11} & T_{12} \\
      T_{21} & T_{22}
    \end{pmatrix}
  \cdot
  \begin{pmatrix}
      M_1\\
      0
    \end{pmatrix}\,,
\end{align}
which is only true for
\begin{equation}
  \label{eq:cond-transfermatrixelement}
  T_{21}\left(n_0,\lambda,a;\beta,\gamma,s_m,s_1,s_2,n_m,\Delta n\right) = 0\,.
\end{equation}
This is the condition from which the complex propagation constants $\beta$ are
found by a two-dimensional root search. To enforce the coalescence into an EP3 the additional conditions 
\begin{equation}
  \label{eq:position-EP-three}
  T_{21} = \frac{\partial T_{21}}{\partial \beta} = \frac{\partial^2 T_{21}}{\partial\beta^2} = 0\,
\end{equation}
have to be obeyed. These additional equations enforce the zero to be
threefold, which is necessary for an EP3. With these equations we are able
to determine $\beta$ as well as the system parameters $\gamma,s_m,s_1$ and
$n_m$ by a six-dimensional root search while we fix $s_2$ and $\Delta n$.

\subsection{Manifestation and verification of an EP3}

\begin{figure}
  \centering
  \includegraphics[width=\linewidth]{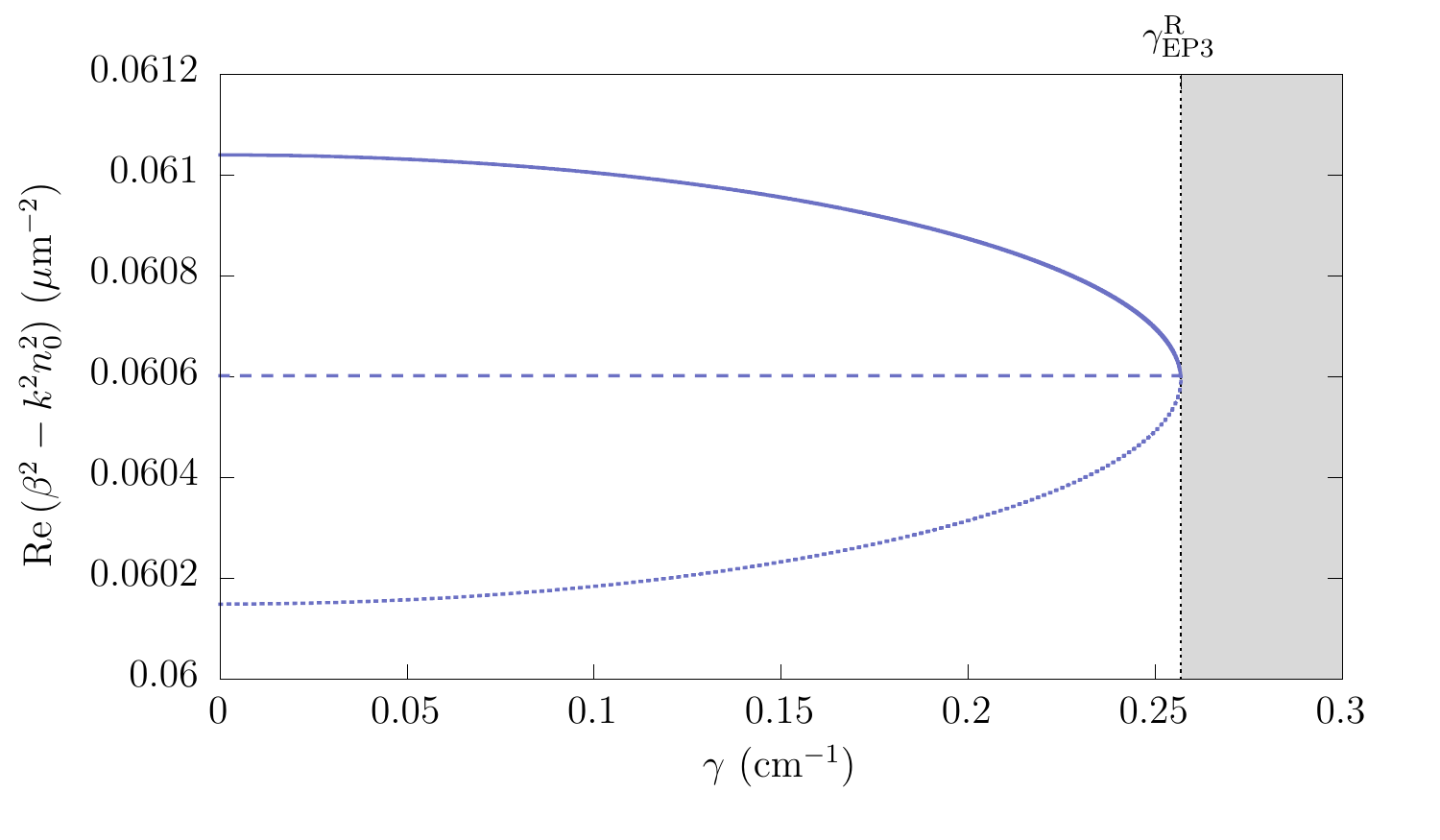}
  \caption{Propagation constants of the three guided modes of the
    wave guide system depicted in Fig.~\ref{fig:system} as a function
    of the non-Hermiticity parameter $\gamma$. As $\gamma$ is
    increased the outer eigenvalues approach each other while the middle
    mode is mostly unaffected by this variation. For
    $\gamma_{\mathrm{EP3}}\approx
    0.2568\,\mathrm{cm}^{-1}$ the eigenmodes coalesce in a third-order
    exceptional point. Beyond this branch point (gray area) the
    propagation constants become complex.}
  \label{fig:full-system-beta-gamma}
\end{figure}

Using the method just described an EP3 is found on the real $\beta$
  axis at
\begin{subequations}
  \allowdisplaybreaks
  \label{eq:EP-params-variable}
  \begin{align}
    \beta_{\mathrm{EP3}} &= 13.37936893005811\,,\\
    \gamma_{\mathrm{EP3}} &= 0.2568441576999367\,,\\
    s_{m}^{\mathrm{EP3}} &= 1.006301260784219\,,\\
    s_{1}^{\mathrm{EP3}} &= 8.983140907622532\,,\\
    n_{m}^{\mathrm{EP3}} &= 1.873188792979378\times 10^{-6}\,,
  \end{align}
\end{subequations}
with the fixed parameters
\begin{eqnarray}
  \label{eq:EP-params-fixed}
  s_{2}^{\mathrm{EP3}} &= 11.0\,\quad\text{and}\quad 
  \Delta n &= 1.3\times 10^{-3}\,.
\end{eqnarray}
The propagation constants of the three guided
modes in the wave guides are plotted in Fig.~\ref{fig:full-system-beta-gamma}
as a function of the non-Hermiticity parameter, with the other system
parameters set to the values according to Eqs.~\eqref{eq:EP-params-variable}
and~\eqref{eq:EP-params-fixed}.
Obviously the first excited state's propagation constant is virtually
independent of the gain-loss parameter gamma. This behavior seems to be
typical for a setup of three coupled wave guides since it also appears for
the idealized delta-functions model \cite{HeissWunner16}. It also appears for
special distributions of the gain and loss in flat band systems
\cite{Ge2015a,Ge2017a}.

It can be seen that increasing $\gamma$ leads to an (inverse) bifurcation
structure of the propagation constants. It is the movement of the
two outer eigenvalues towards each other with an essentially constant middle
value into the third-order exceptional point at $\gamma_{\mathrm{EP3}}$. Beyond
this point (gray area) the propagation constants become complex. Thus we
confirm the findings for two coupled \PT-symmetric wave guides
in~\cite{Klaiman08}, i.e., one may study the exceptional point by varying only
a \emph{single parameter}. This is in contrast to the idealized delta-functions
model \cite{HeissWunner16}, in which two parameters have to be varied in steps
to reach the EP3 starting from $\gamma = 0$.

For this, however, it is necessary to adjust the system parameters exactly
according to Eqs.~\eqref{eq:EP-params-variable} and~\eqref{eq:EP-params-fixed}
to end up in an EP3 within a numerical simulation. Deviations from these
values will lead to a coalescence of merely two modes. At this point we
encounter the perhaps most difficult part in an experimental realization --
the exceeding sensitivity to changes in the setup of the system parameters.

\begin{figure}
  \centering
  \includegraphics[width=\linewidth]{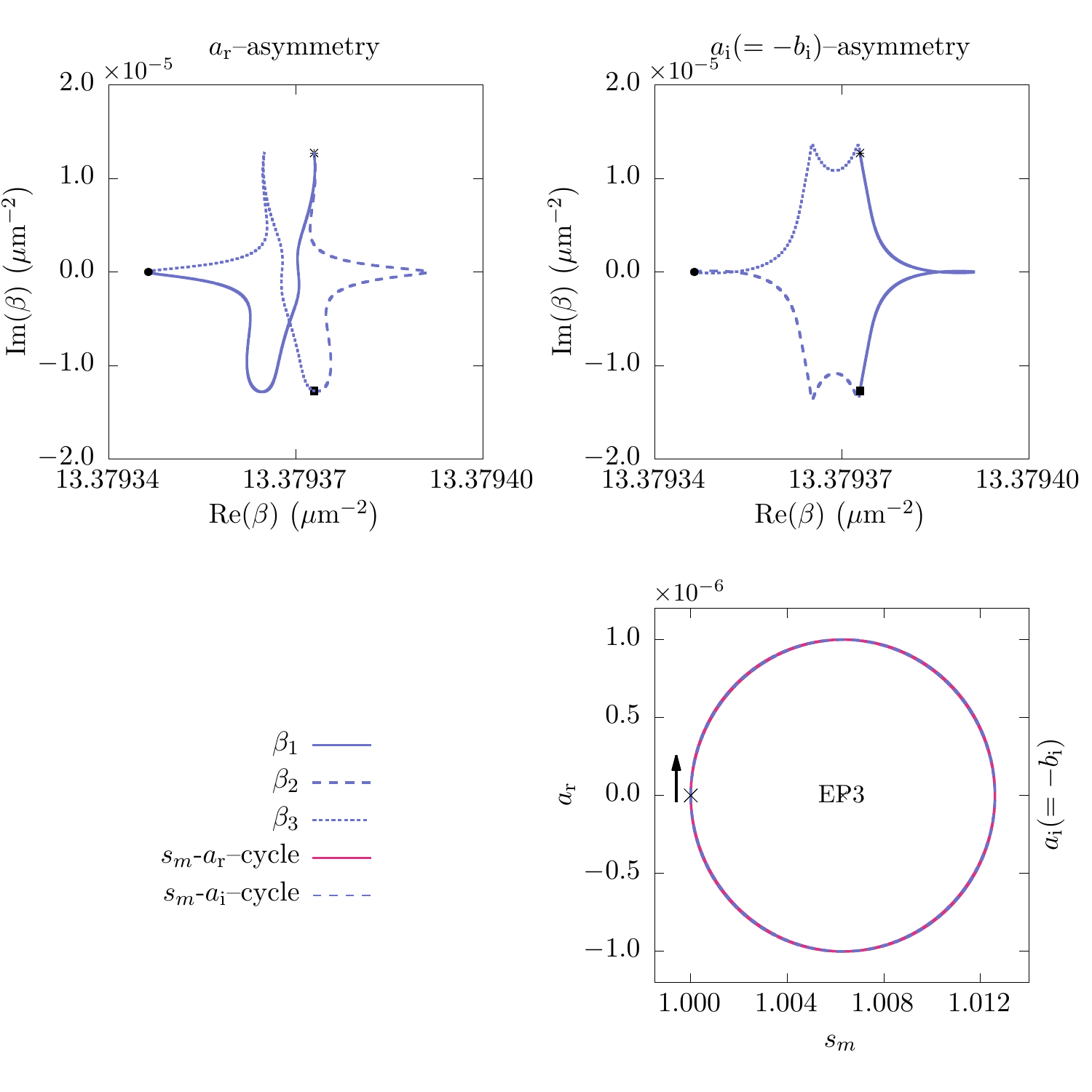}
  \caption{Verification of the EP3 in
    Fig.~\ref{fig:full-system-beta-gamma} by encircling it in the
    space of asymmetry parameters (breaking the systems \PT\ symmetry
    in either the real or imaginary part of the refractive index) and
    distance (represented by $s_m$). The upper panel shows the
    characteristic threefold permutation if one encircles the
    EP3 in the $s_m$-$a_{\mathrm{r}}$--space (left) or in the
    $s_m$-$a_\ii~(= - b_\ii)$--space (right). The starting points of the
    respective eigenvalues are depicted by specific symbols.
    They are the end points of the paths of another eigenvalue.
    Thus, one can see that the disturbed ground state branch (left-most
    starting point) ends at the starting point of one of the excited states.
    The path starting there ends at the starting point of the second of the
    excited states. Finally, the path of the latter ends at the starting
    point of the disturbed ground state.
    The corresponding curves shown on the bottom right are parametrized
    according to Eq.~\eqref{eq:paramerization-realistic}, where the
    loop is performed clockwise for both circles.}
  \label{fig:full-system-cycles}
\end{figure}

We verify the expected properties of the EP3 by encircling the branch point.
We introduce asymmetry parameters breaking the underlying \PT\ symmetry by
adding $a = a_{\mathrm{r}} +\ii a_\ii$ to the refractive index of the left wave
guide and $b = b_{\mathrm{r}} + \ii b_\ii$ to the right one, which changes the
parameters $q_l$ and $q_r$ from Eqs.~\eqref{eq:wavefuncs-params-b}
and~\eqref{eq:wavefuncs-params-d}, viz.
\begin{subequations}
  \label{eq:asymmetric-wavefunc-params}
  \begin{align}
    \tilde{q}_l^2 &= -\beta^2 + k^2\Bigl[n_0+\Delta n + {a_{\mathrm{r}}} 
    - \ii\Bigl(\frac{\lambda}{2\pi}\gamma + {a_\ii}\Bigr)\Bigr]^2\,,\\
    \tilde{q}_r^2 &= -\beta^2 + k^2\Bigl[n_0+\Delta n + {b_{\mathrm{r}}} 
    + \ii\Bigl(\frac{\lambda}{2\pi}\gamma + {b_\ii}\Bigr)\Bigr]^2\,.
  \end{align}
\end{subequations}
We break the \PT\ symmetry in either the real or the imaginary part of the
refractive index. We perform the loop in the space of this asymmetry and the
distance between the wave guides. The distance can be varied
with $s_m$ whence the loop can be parametrized as
\begin{eqnarray}
  \label{eq:paramerization-realistic}
  \left(\begin{array}{c}
      s_m\\
      a_{\mathrm{r}}
    \end{array}\right)
  &=
  \left(\begin{array}{c}
      s_{m}^{\mathrm{EP3}} + \left(1-s_{m}^{\mathrm{EP3}}\right)\cos\varphi\\
      10^{-6}\sin\varphi
    \end{array}\right)
\end{eqnarray}
with $\varphi\in\left[0,2\pi\right]$ for an asymmetry in the real part
as only the refractive index of the left wave guide is varied
($b_{\mathrm{r}} = a_\ii = b_\ii = 0$). A similar parametrization can
be used for an asymmetric variation of the imaginary part, i.e., $a_\ii
= -b_\ii$. Both situations are depicted in
Fig.~\ref{fig:full-system-cycles}.
The characteristic threefold permutation of the propagation constants
becomes obvious in both cases while the circle for the
$a_\ii/-b_\ii$--asymmetry shows higher symmetry compared to the loop
performed in the $s_m$-$a_{\mathrm{r}}$--space.

Note that the numerical precision achieved in this work as shown in
Eq.~\eqref{eq:EP-params-variable} is not realizable in an experiment. However,
as the EP3 splits into two EP2s under a generic perturbation, the threefold
permutation remains unchanged when both EP2s are encircled, i.e. for small
perturbations the EP3-signature persists. We checked that this is true for
deviations from the EP3 on the order of the circle radii used above.

\subsection{Stationary eigenmodes and power distribution}

\begin{figure}
  \centering
  \includegraphics[width=.9\linewidth]{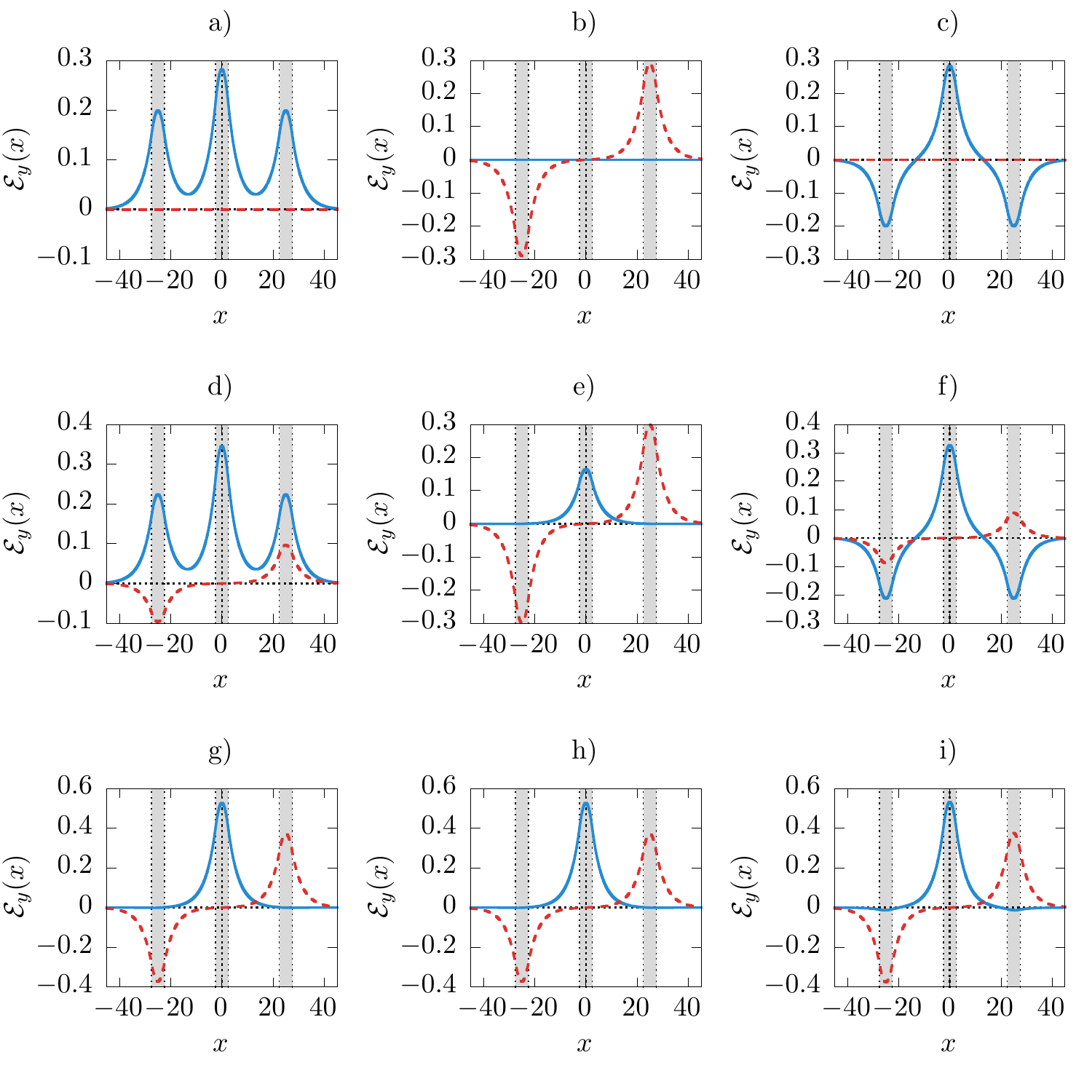}
  \caption{C-normalized stationary modes of the wave guide system for
    increasing non-Hermiticity parameter (from top to bottom)
    $\gamma=0.0\,\mathrm{cm}^{-1}$ (\textbf{a)} - \textbf{c)}),
    $\gamma=0.1\,\mathrm{cm}^{-1}$ (\textbf{d)} - \textbf{f)}),
    $\gamma=0.256802\,\mathrm{cm}^{-1}$ (\textbf{g)} -
    \textbf{i)}). The real part of the modes is illustrated by blue
    solid lines and the imaginary part by red dashed lines. From left to right
    there are shown the ground state mode and the two excited modes. The gray
    shaded areas represent the wave guides' positions.}
  \label{fig:full-system-wavefunctions}
\end{figure}

For the analysis of the stationary eigenmodes of the wave guide system
the corresponding coefficients of Eq.~\eqref{eq:eigenmodes} have to be
calculated first. We recall that physically meaningful modes occur with
$\tilde{\mathcal{E}}_y(x)\to 0$ for $x\to\pm\infty$, i.e., $A_2$ and $M_2$ must
vanish. One of the coefficients can be chosen freely and without loss of
generality we fix $M_1 = 1$. Consequently we obtain an additional overall
phase $\varphi_0$
\begin{align}
  \label{eq:global-phase}
  \varphi_0 &= \arctan\left(\frac{\Im [\tilde{\mathcal{E}}_y(0)]}{\Re [\tilde{\mathcal{E}}_y(0)]}\right)\nonumber \\
  &= \arctan\left(\frac{\Im (F) + \Im (G)}{\Re (F) + \Re (G)}\right)\,,
\end{align}
which has to be compensated to ensure exact \PT\ symmetry. Because of the
non-Hermiticity we have to use the \emph{c norm} \cite{Moiseyev} $N_c$
defined via
\begin{equation}
  \label{eq:c-norm}
  \frac{1}{N_c^2}\int_{-\infty}^{\infty}\,\tilde{\mathcal{E}}_y^2(x)\,\mathrm{d}x
  = 1\, ,
\end{equation}
which, for the underlying \PT\ symmetry, can easily be calculated from the
real and imaginary parts of $\mathcal{E}$ since the real part is an even
function of $x$, whereas the imaginary part is odd. Consequently the integral
splits into the difference of the $L^2$ norms taken separately. Thus, the
stationary modes illustrated in Fig.~\ref{fig:full-system-wavefunctions}
for some values of $\gamma$ are calculated as
\begin{equation}
  \label{eq:modes}
  \mathcal{E}_y(x) = \frac{\ee^{-\ii\varphi_0}\tilde{\mathcal{E}}_y(x)}{N_c}\,.
\end{equation}
The modes depicted correspond to a system configuration according to
Eqs.~\eqref{eq:EP-params-variable} and~\eqref{eq:EP-params-fixed}. In
line with the system's \PT\ symmetry the real part of the modes is
symmetric and the imaginary part is antisymmetric. With increasing
$\gamma$ the imaginary part of the ground state mode and second
excited mode grows while it is the real part that increases for the
first excited mode. Close to the EP3 (bottom panel) we obtain the
expected self-orthogonality phenomenon as the modes become essentially
equal.

\begin{figure}
  \centering
  \includegraphics[width=\linewidth]{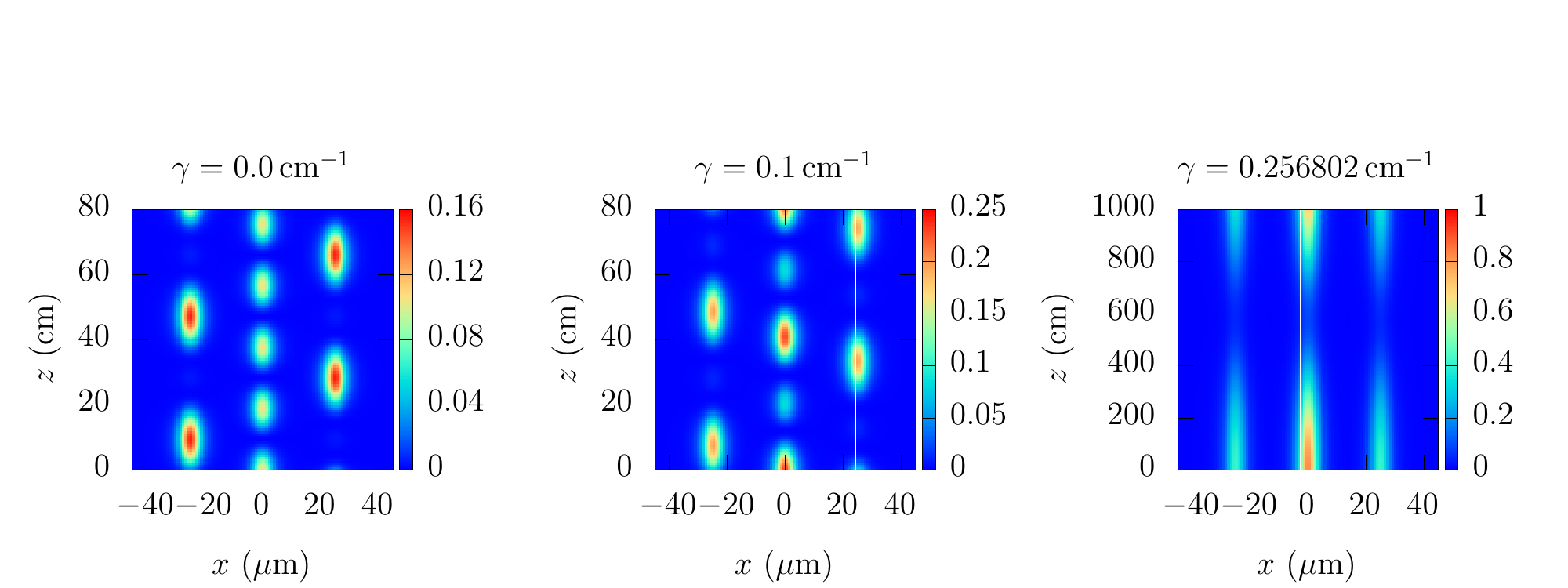}
  \caption{The power distribution for the propagating total field
    consisting of the three guided modes, see
    Eq.~\eqref{eq:power-distribution}, for three values of the
    non-Hermiticity $\gamma$. With increasing values of $\gamma$ the
    corresponding beat length also increases. In addition an obvious
    rise in the intensity can be observed.}
  \label{fig:full-system-dynamics}
\end{figure}

The progression of the propagation constants on the real axis towards
the branch point according to Fig.~\ref{fig:full-system-beta-gamma} as
well as the self-orthogonality phenomenon can be visualized
experimentally by observing the beat length $L = 2\pi/\Delta\beta$,
where $\Delta\beta$ is the difference between two modes, of the power spectrum
for the \PT-symmetric wave guide system. This can be observed
for a non-stationary state. The power distribution
\begin{equation}
  \label{eq:power-distribution}
  \left|E_y(x,z)\right|^2 = \left|\frac{1}{\sqrt{3}}\sum_{i=1}^3\mathcal{E}_i(x)\ee^{-\ii\beta_i z}\right|^2
\end{equation}
is taken, and displayed in Fig.~\ref{fig:full-system-dynamics} for three
different values of $\gamma$.
With increasing $\gamma$ the beat length also increases, which is a direct
consequence of the movement of the propagation constants towards each other
($\Delta\beta$ becomes smaller). In the vicinity of the exceptional point the 
power spectrum no longer oscillates between the wave guides but rather pulses
in all three wave guides simultaneously. Note the different length scales for
the direction of propagation (i.e., $z$ axis). As the branch point is
approached, i.e., $\Delta\beta\approx 0$ the beat length goes to infinity.

Furthermore we observe an increasing intensity of the power field for
increasing values of the non-Hermiticity (see the corresponding color
bars). This phenomenon is a consequence of the vanishing c norm when the
branch point is approached. We note that the results shown in
Fig.~\ref{fig:full-system-wavefunctions} and
Fig.~\ref{fig:full-system-dynamics} for extended wave guides
are in line with those of the simple three delta-functions model discussed in 
\cite{HeissWunner16}, confirming the validity of that model.
 
\section{Further exceptional points in parameter space in the vicinity of the EP3}
\label{sec:EP2s}

In this section we address an aspect associated with higher-order EPs that
is related to the high parameter sensitivity of the eigenmodes in the vicinity
of the EP3. It is a phenomenon that has so far attracted little attention but
an awareness appears to be of utmost importance in an expected experimental
confirmation. As is qualitatively discussed in \cite{HeissChiral,Graefe12} a
perturbation by only one of the parameters that were chosen to invoke the
third-root branch point infers three eigenvalues to pop out in the energy
plane from the EP3. In turn, the EP3 can be seen as a coalescence of two EP2s
as the three eigenvalues -- obtained from this perturbation -- are still
analytically connected. In fact, searching for singularities using some other
parameter one finds two EP2s that sprout from the original EP3. Yet another
parameter could then be used to force a coalescence of the two EP2s into a new
and therefore shifted EP3.

This generic pattern turns out to be crucial for the identification of the
EP3 via parameter space loops in our system. In the space of the physical 
parameters at hand curves of second-order and third-order exceptional points
are found. These have a decisive effect on the permutation behavior of the
modes.

\begin{figure}[]
  \centering
  \includegraphics[width=\linewidth]{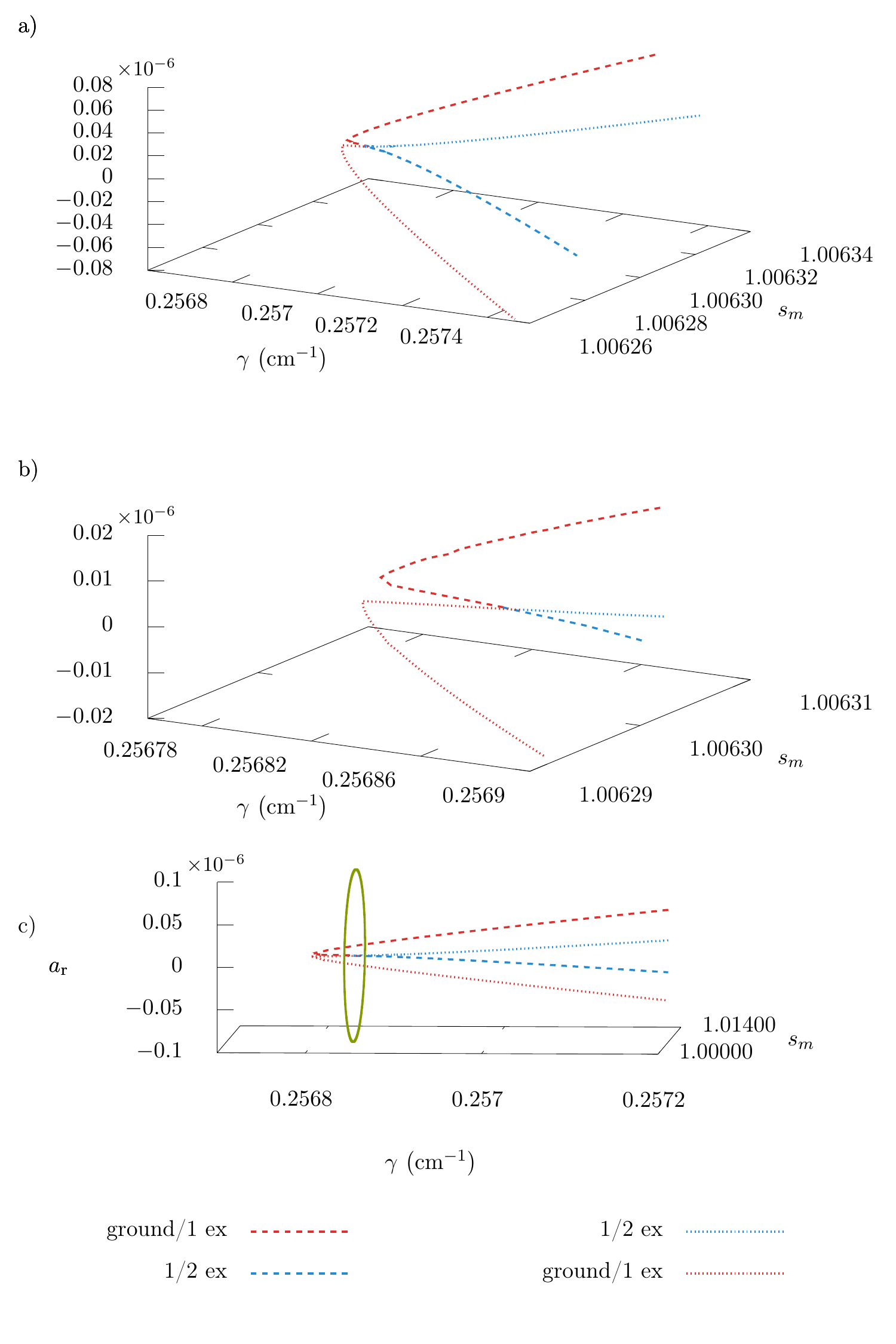}
  \caption{\textbf{a)} Curves denoting the positions in the three-dimensional
    parameter space $\left(\gamma,s_m,a_{\mathrm{r}}\right)$ where EP2s are
    found for the wave guide system depicted in Fig.~\ref{fig:system}. The
    parameters $s_1^{\mathrm{EP3}}$, $s_2^{\mathrm{EP3}}$, $n_m^{\mathrm{EP3}}$,
    and $\Delta n$ are held fixed and for every $\gamma$ the values of
    $\Re (\beta), \Im (\beta),s_m,a_{\mathrm{r}}$ are determined in a
    four-dimensional root search such that
    Eq.~\eqref{eq:cond-ep-two-full-system} is fulfilled.
    Branches of EP2s connecting either ground state and first excited
    mode or first and second excited mode  sprout out from the
    EP3.
    \textbf{b)} Magnification of the space around the EP3 from which
    all lines originate. \textbf{c)} The circle in the parameter
    space  (solid green
    line) in Fig.~\ref{fig:full-system-cycles} circumscribes the EP3 and two
    EP2s formed by the branches of the ground state and the first excited
    state.}
  \label{fig:EP2-real-system-ar}
\end{figure}

\begin{figure}[]
  \centering
  \includegraphics[width=\linewidth]{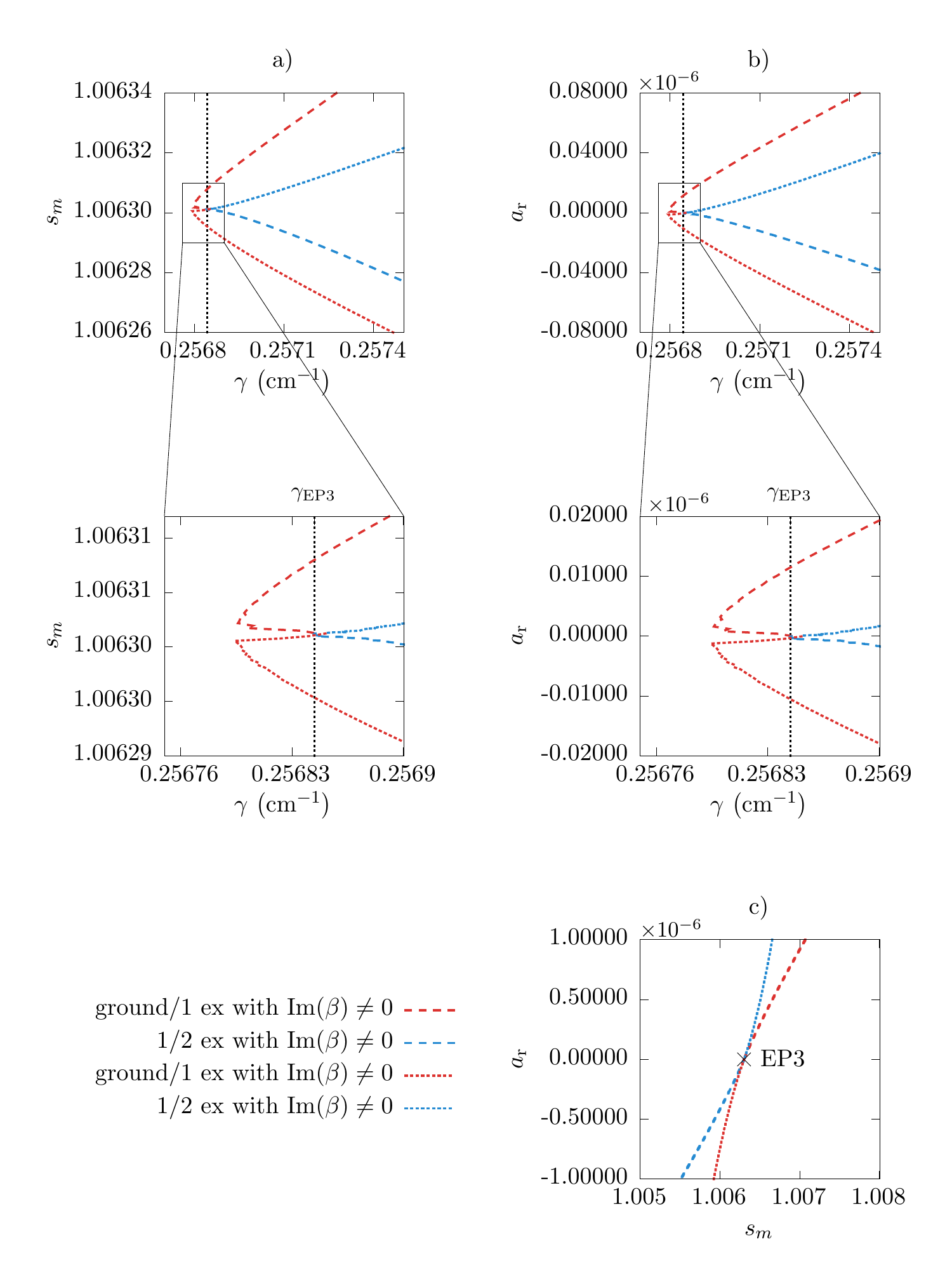}
  \caption{Projections of the two-dimensional curves of EP2s from
      Fig.~\ref{fig:EP2-real-system-ar} on the \textbf{a)} $\gamma$-$s_m$
      \textbf{b)} $\gamma$-$a_\mathrm{r}$ and \textbf{c)} $s_m$-$a_\mathrm{r}$
      planes.
  }
  \label{fig:EP2-real-system-ar-2dim}
\end{figure}

To discover curves of EP2s in the system, and to clarify the points raised
above we use the following condition similar to
Eq.~\eqref{eq:position-EP-three},
\begin{equation}
  \label{eq:cond-ep-two-full-system}
  T_{21} = \frac{\partial T_{21}}{\partial\beta} = 0\,.
\end{equation}
As $T_{21}$ and its first derivative are complex valued functions, these
equations give us four conditions that have to be fulfilled. 
Results are illustrated in 
Fig.~\ref{fig:EP2-real-system-ar}
for a configuration of the system close to the EP3 given by
Eqs.~\eqref{eq:EP-params-variable} and~\eqref{eq:EP-params-fixed}. While
$s_1$, $s_2$, $n_m$, and $\Delta n$ are held fixed to their values at the EP3,
$\gamma$ is varied in the range shown in the figures. For each value of
$\gamma$ the parameters $\Re (\beta), \Im (\beta),s_m,a_{\mathrm{r}}$ are
determined in a four-dimensional root search.
Projections of these lines on the two-dimensional parameter planes
  are shown in Fig.~\ref{fig:EP2-real-system-ar-2dim}.

While it is true that the threefold permutation identified in
Fig.~\ref{fig:full-system-cycles} clearly indicates
the topological character of an EP3 one must keep in mind that
the path around the EP3 circles in addition two second-order
exceptional points, each of them formed by the ground and the first
excited state. They belong to the red dashed and dotted lines.
This explains why we do not find simple circles in
Fig.~\ref{fig:full-system-cycles} but rather the twisted curves that are
caused by the presence of the EP2s (see also Fig.~7 in \cite{Menke2016a}
in a similar context).
The effect of the two exceptional points included in the
encircling is such that it does not affect the threefold permutation,
i.e.~the EP3 remains visible. In fact, both EP2s share the same sheet.
It guarantees that the
threefold permutation is not disturbed by their combined action. Of course
the inclusion of the EP2s can be avoided altogether with a smaller circle. However, in
an experiment it would be a rather laborious task to find and characterize
all of the exceptional points thus avoiding the inclusion of unwanted EP2s.

The situation is different if $\gamma$ is chosen as one of the parameters for
the circle. As can be extracted from
Figs.~\ref{fig:EP2-real-system-ar-2dim} (a) and (b) the
curves of EP2s only appear for non-zero values of the asymmetry parameter
$a_{\mathrm{r}}$, which implies that the propagation constants become complex.
At the position of the EP3, different EP2 lines originate. Along the blue
dashed and dotted lines there are EP2s connecting the two excited modes
that differ only in the signs of the corresponding imaginary parts of the
propagation constants $\beta$. At the EP3 these imaginary parts vanish.
The situation is similar along the red dashed and dotted lines, where
the ground state mode and first excited mode are connected by an EP2.
Since  for $a_{\mathrm{r}}=0$ there exist no second-order exceptional points in
the $\gamma$-$s_m$ plane nor in the $\gamma$-$a_{\mathrm{r}}$ plane we should,
in principle be able to verify the EP3 by encircling. Yet it turns out that we
observe EP2-like signatures.

\begin{figure}
  \centering
  \includegraphics[width=\linewidth]{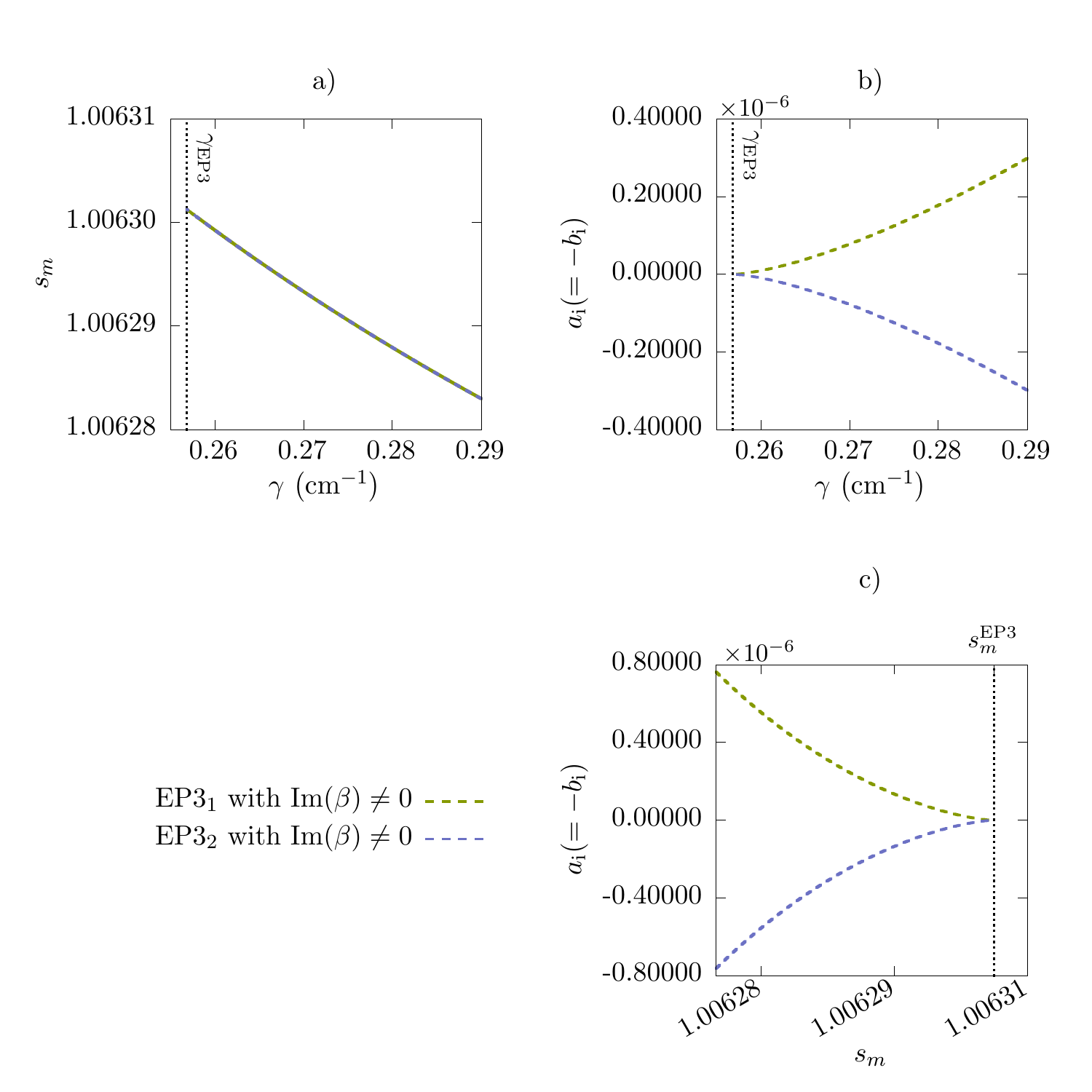}
  \caption{Curves denoting the positions in three-dimensional parameter space
    $\left(\gamma,s_m,a_\ii=-b_\ii\right)$ where EP3s are found for the wave
    guide system depicted in Fig.~\ref{fig:system}. For every $\gamma$
    $\Re (\beta),\Im (\beta),s_m,a_\ii=-b_\ii$ are determined in a
    four-dimensional root search such that
    Eq.~\eqref{eq:cond-ep-two-full-system} is fulfilled while
    the other parameters of Eqs.~\eqref{eq:EP-params-variable}
    and~\eqref{eq:EP-params-fixed} are held fixed.
    The figures contain the projections on the respective parameter
    planes. EP3s only exist in regions where $a_\ii=-b_\ii$ assumes non-zero
    values. The distributions differ in so far as they belong to
    different values of $a_\ii=-b_\ii$, which is also noticeable in mirrored
    imaginary parts of the propagation constants (not shown).}
  \label{fig:EP2-real-system-ai-new}
\end{figure}

If we allow for $a_\ii=-b_\ii\neq 0$ we obtain the results shown in
Fig.~\ref{fig:EP2-real-system-ai-new}. Instead of EP2s, signatures of EP3s can
clearly be discerned (dashed lines). The corresponding propagation
constants have a non-vanishing imaginary part as expected due to the broken
$\mathcal{PT}$-symmetry. Again it should be possible to observe the EP3
signature in the $\gamma$-$s_m$ plane as well as in the $\gamma$-$a_\ii/b_\ii$
plane, respectively. However, only an EP2 signature is found. As in the example
in the previous paragraph it could be possible that the EP3 interacts in such
a way that the result is an EP2 signature \cite{Graefe12,Heiss2016}.

Thus there are different curves of EP2s and EP3s associated with the EP3
of the actual system.
In this model there are specific parameter planes that are free from any EP2,
yet they  cannot be used to show the existence of the EP3 simply by encircling. 
Thus, the EP2 and EP3 lines
have to be taken into account when the EP3 is supposed to be detected via its
permutation behavior.

\section{Conclusion and Outlook}
\label{sec:summary}

A system of three coupled \PT-symmetric wave guides can serve as a promising
setup for an experimental verification of third-order exceptional points.
Within an experimentally realizable parameter range for the system we have
shown that the EP3 can be determined by simply varying a
\emph{single parameter} once the other parameters have been properly tuned.
In our approach the non-Hermiticity parameter $\gamma$ is varied. The proper
tuning of the other parameters appears to be the most challenging part in an
experiment as even in numerical calculations, where the necessary high
precision can be achieved, the task of finding the EP3 is rather demanding.

We feel that in a measurement of the power distributions of the total field
for \PT-symmetric wave guides a direct visualization of the progression of
the propagation constants towards the branch point can be obtained. It can be
discerned by the increasing beat length when the EP3 is approached. In
addition, using an appropriate encircling around the assumed position of the
branch point it is possible to verify the threefold state exchange without even
knowing the point's exact position. Our numerical study can guide the
approximate localization of the EP3 in an experimental setup.

Related to the verification of an EP3 by observing a threefold state exchange
when encircling it in a suitably chosen parameter plane we have also shown
that the branch point has further satellites of branches of EP2s
or EP3s. From this we can extract a possible explanation for the complicated
exchange behavior. They influence the permutation behavior and
complicate the verification of the EP3 via its characteristic threefold
state exchange. Thus, the beat length mentioned above might be the best choice
for an experimental proof.

In a next step we will extend this one-dimensional optical system to a
three-dimensional quantum mechanical one in terms of a Bose-Einstein
condensate in a triple-well potential. This way we are going to propose a
further, now really quantum mechanical, \PT-symmetric system for the
verification of a third-order exceptional point.

GW and WDH gratefully acknowledge support from the National Institute for
Theoretical Physics (NITheP), Western Cape, South Africa. GW expresses his
gratitude to the Department of Physics of the University of Stellenbosch
where this paper was finalized.

\end{document}